\documentclass[a4paper,11pt]{article}
\pdfoutput=1 
\usepackage{jcappub}
\usepackage{amsmath}
\usepackage{graphicx}

\usepackage{geometry}
\usepackage{xspace}
\usepackage{pdflscape}

\makeatletter
\gdef\@fpheader{}
\g@addto@macro\bfseries{\boldmath}
\makeatother

\usepackage[english]{babel}
\usepackage[utf8]{inputenc}
\usepackage{pdflscape}
\usepackage{jcappub}
\usepackage{enumerate}
\usepackage{amsbsy}
\usepackage{amsmath} 
\usepackage{graphics}
\usepackage{mathrsfs}
\usepackage{wrapfig}
\usepackage{mathtools}

\usepackage{amsfonts}
\usepackage{pstricks}
\usepackage{color}
\usepackage{setspace}

\newcommand{\Eq}[1]{Eq.~(\ref{#1})}
\newcommand{\Eqs}[1]{Eqs.~(\ref{#1})}

\newcommand{\Ref}[1]{Ref.~{\cite{#1}}}
\newcommand{\Refs}[1]{Refs.~{\cite{#1}}}

\newcommand{\bea}{\begin{eqnarray}} \newcommand{\eea}{\end{eqnarray}}
\newcommand{\el}{\nonumber \\}
\newcommand{\re}[1]{(\ref{#1})}

\newcommand{\pat}{\partial}
\newcommand{\abs}[1]{|#1|}
\renewcommand{\sec}[1]{section \ref{#1}}
\newcommand{\fig}[1]{figure \ref{#1}}

\newcommand{\para}{\paragraph}

\renewcommand{\a}{\alpha}
\renewcommand{\b}{\beta}
\renewcommand{\c}{\gamma}
\renewcommand{\d}{\delta}

\renewcommand{\l}{\lambda}

\newcommand{\ha}{\frac{1}{2}}

\newcommand{\rmd}{\mathrm{d}}

\newcommand{\ie}{i.e.\ }

\newcommand{\av}[1]{\langle{#1}\rangle}

\newcommand{\f}{\frac}
\newcommand{\mA}{\mathcal{A}}
\newcommand{\Mpl}{M_{{}_{\mathrm{Pl}}}}
\newcommand{\rme}{{\mathrm{end}}}
\newcommand{\rmb}{{\mathrm{beg}}}
\newcommand{\reh}{{\mathrm{reh}}}

\newcommand{\se}{\sigma_{k}}
\newcommand{\fmin}{f_{\mathrm{min}}}

\def\beq{\begin{equation}}
\def\eeq{\end{equation}}
\def\baq{\begin{eqnarray}}
\def\eaq{\end{eqnarray}}

\title{Narrowing the window of inflationary magnetogenesis}

\author[a]{Tommi Markkanen,}

\author[b]{Sami Nurmi,}

\author[c]{Syksy R\"{a}s\"{a}nen}

\author[d]{and Vincent Vennin}

\affiliation[a]{Department of Physics, King’s College London \\
Strand, London WC2R 2LS, United Kingdom}

\affiliation[b]{Department of Physics, University of Jyv\"{a}skyl\"{a} \\
P.O. Box 35 (YFL), FI-40014 University of Jyv\"{a}skyl\"{a}, Finland}

\affiliation[c]{Department of Physics
and Helsinki Institute of Physics, University of Helsinki  \\
P.O. Box 64, FIN-00014 University of Helsinki, Finland \\
and Department of Physics, Kobe University, Kobe 657-8501, Japan}

\affiliation[d]{Institute of Cosmology \& Gravitation, University of Portsmouth \\
Dennis Sciama Building, Burnaby Road, Portsmouth, PO1 3FX, United Kingdom}

\emailAdd{tommi.markkanen@kcl.ac.uk}
\emailAdd{sami.t.nurmi@jyu.fi}
\emailAdd{syksy.rasanen@iki.fi}
\emailAdd{vincent.vennin@port.ac.uk}

\abstract{We consider inflationary magnetogenesis where the conformal symmetry is broken by the term $f^2(\phi) F_{\a\b} F^{\a\b}$. We assume that the magnetic field power spectrum today between 0.1 and $10^4$ Mpc is a power law, with upper and lower limits from observation. This fixes $f$ to be close to a power law in conformal time in the window during inflation when the modes observed today are generated. In contrast to previous work, we do not make any assumptions about the form of $f$ outside these scales. We cover all possible reheating histories, described by an average equation of state $-1/3 <\bar{w} <1$. Requiring that strong coupling and large backreaction are avoided both at the background and perturbative level, we find the bound $\delta_{B_0} < 5 \times10^{-15} \left( \frac{r}{0.07} \right)^{1/2} \kappa \mathrm{G}$ for the magnetic field generated by inflation, where $r$ is the tensor-to-scalar ratio and $\kappa$ is a constant related to the form of $f$. This estimate has an uncertainty of one order of magnitude related to our approximations. The parameter $\kappa$ is $<100$, and values $\gtrsim1$ require a highly fine-tuned form of $f$; typical values are orders of magnitude smaller.}

\begin{document}

\begin{flushleft}
	\hfill		 HIP-2017-04/TH \\
	\hfill		 KCL-PH-TH/2017-15\\
	\hfill		 KOBE-COSMO-17-04
\end{flushleft}

\maketitle
  
\setcounter{tocdepth}{2}

\setcounter{secnumdepth}{3}

\section{Introduction} \label{sec:intro}

\para{Cosmic magnetogenesis.}

There seem to be cosmic magnetic fields from galactic scales all the way up to the largest observable scale of $10^4$ Mpc \cite{Durrer:2013pga}. Magnetic fields in galaxies and clusters are of the order $10^{-5}$ G to $10^{-6}$ G, while on cosmological scales their amplitude is poorly known, with an upper limit of $10^{-9}$ G and there seems to be a conservative lower limit of $10^{-17}$ G.
Galactic fields are likely generated from much smaller seed fields by the dynamo mechanism \cite{Durrer:2013pga, Pakmor:2013rqa}.
The origin of the seed fields, as well as magnetic fields with large correlation lengths, unaffected by magnetohydrodynamic processes, remains unexplained.

A natural possibility to obtain fields with large correlation lengths is to generate them during inflation. In inflation, quantum fluctuations of scalar perturbations are amplified and their amplitude freezes out as the wavelength is stretched above the Hubble scale. As electromagnetic fields are conformally invariant, expansion does not have a similar effect on them. Inflationary magnetogenesis therefore requires breaking the conformal symmetry of electromagnetism \cite{Turner:1987bw}. Possibly the simplest way of doing so (apart from coupling the electromagnetic field strength to the Riemann tensor, which does not give a large enough amplitude~\cite{Turner:1987bw}) is to couple the electromagnetic field to a scalar field $\phi$, possibly the inflaton, via the term $f(\phi)^2 F_{\a\b} F^{\a\b}$, where $F_{\a\b}$ is the field strength \cite{Ratra:1991bn}. Canonically normalising the kinetic term then leads to a field-dependent modification of the electric coupling, and keeping it perturbative constrains the range of validity of any study that does not take non-perturbative QED into account \cite{Ratra:1991bn}. The energy density of the electromagnetic field also must not be so large as to interfere with inflation and magnetogenesis \cite{Demozzi:2009fu, Fujita:2012rb, Ferreira:2013sqa} (though having a significant fraction of the energy density in the electromagnetic field does not necessarily prevent inflation \cite{Kanno:2009ei, Kanno:2008gn,Dimopoulos:2010xq,Dimopoulos:2011ym, Karciauskas:2016pxn}).
Moreover, even if the electromagnetic field is subdominant for the background, it is important to check that its perturbations do not spoil the success of the inflationary generation of a Gaussian spectrum of nearly scale-invariant adiabatic perturbations \cite{Giovannini:2007aq, Suyama:2012wh, Giovannini:2013rme, Ringeval:2013hfa, Bonvin:2013tba, Fujita:2014sna, Ferreira:2014hma, Ade:2015cva}. As the magnetic field has vanishing background and Gaussian perturbations, its energy density, quadratic in the field, induces non-Gaussian perturbations \cite{Caprini:2009vk, Caldwell:2011ra, Motta:2012rn, Jain:2012ga, Barnaby:2012tk, Lyth:2013kah, Fujita:2013pgp, Nurmi:2013gpa, Ferreira:2014hma}, which are strongly constrained by observations.
Much of the work has assumed that either the coupling function $f$ \cite{Demozzi:2009fu, Bonvin:2013tba, Ferreira:2014hma} or some quantities more directly related to the magnetic field \cite{Suyama:2012wh, Ringeval:2013hfa, Fujita:2014sna} is a power law, though \cite{Ferreira:2013sqa} considered a more involved form.

We extend previous work by only assuming that the magnetic field spectrum today is a power law in the observable region (which leads to $f$ essentially being a power law during the era in inflation when the observed magnetic fields are generated), without assumptions about the shape outside of that region. We only consider non-helical magnetic fields.

In \sec{sec:iruv} we describe our setup and the asymptotic matching method used to obtain super-Hubble solutions, and test it in a case where the exact solution is known. In \sec{sec:con} we compare the theoretical power spectrum against observations, taking into account the constraints discussed above. In \sec{sec:conc} we compare to previous work and summarise our results. In appendix \ref{sec:app} we show that a power-law magnetic power spectrum leads to a power-law form for $f$ in the observable window.

\section{The magnetogenesis setup} \label{sec:iruv}

\subsection{Non-conformal coupling} \label{sec:setup}

\para{The action and the equation of motion.}

We follow the notations of \Ref{Bamba:2003av, Bamba:2004cu, Martin:2007ue}, where more details can be found about the basic formalism. We consider the action
\bea \label{action} 
  S \left[ \phi, A_{\mu} \right] &=& - \frac14\int {\rm d}^4x \sqrt{-g} f^2(\phi) F_{\mu\nu}F^{\mu\nu} + S_{\mathrm{other}} \, ,
\eea
where $F_{\mu\nu}=\nabla_\mu A_\nu-\nabla_\nu A_\mu$ is the electromagnetic field tensor, $\phi$ is a scalar field (which may or may not be the inflaton), $f$ is a so far unspecified function, and $S_{\mathrm{other}}$ contains all other terms which, we assume, do not break the conformal invariance of $A_\mu$ at tree level. We neglect any other interaction terms between $A_\mu$ and other fields. If the interaction energy density associated with such terms is negative, this could weaken our constraints based on demanding that the electromagnetic energy density is not too large, discussed in \sec{sec:BR}. We consider only non-helical magnetic fields. Variation of \Eq{action} gives the equation of motion for $A_\mu$,
\begin{eqnarray} \label{eom}
  \partial _{\mu}\left[\sqrt{-g}f^2\left(\phi \right)F^{\mu \nu }\right] &=& 0 \ .
\end{eqnarray} 
In a spatially flat Friedmann-Lema\^itre-Robertson-Walker universe with the metric
\begin{eqnarray}
\label{metric}
  {\rm d}s^2 = a^2(\eta)\left(-{\rm d}\eta^2 + {\rm d}{\boldsymbol{x}}^2\right) 
\end{eqnarray}
and in the Coulomb gauge, where $A_0=0$ and $\partial_i A^i=0$, \Eq{eom} reads
\begin{eqnarray}
\label{eomcosmic}
  A_i'' + 2\frac{f'}{f}A_i' - \d^{jk} \pat_j \pat_k A_i &=& 0 \, ,
\end{eqnarray}
where prime denotes derivative with respect to conformal time $\eta$.

The quantised field $A_i(t,\boldsymbol{x})$ can then be written as
\begin{eqnarray} \label{FourierA}
  A_i(\eta ,\boldsymbol{x}) &=&
\int \frac{{\rm d}^3 k}{(2\pi)^{3/2}}\sum_{\lambda =1}^2
\epsilon_{i \lambda }(\boldsymbol{k})\biggl[
b_{\lambda} (\boldsymbol{k}) A(\eta,k){\rm e}^{i \boldsymbol{k} \cdot
  \boldsymbol{x} } 
+ b_{\lambda }^{\dagger}(\boldsymbol{k}) {A}^*(\eta ,k){\rm e}^{-i
  \boldsymbol{k} \cdot \boldsymbol{x}} \biggr] \, , 
\end{eqnarray}
where $\boldsymbol{k}$ is the comoving wavenumber, with the completeness and orthogonality relations
\begin{equation}
\label{complet}
\sum _{\lambda =1}^2\epsilon ^{i}_{\lambda } (\boldsymbol{k})
\epsilon _{j \lambda }(\boldsymbol{k})
=\delta ^i{}_j - \hat k^i \hat k_j \, ,\qquad \epsilon_\lambda^{i}(\boldsymbol{k}){\epsilon_\lambda}_{i}(\boldsymbol{k})=1 \, ,
\end{equation}
where $\hat{k}^i\equiv k^i/\sqrt{k^j k_j}$, and $b_{\lambda}(\boldsymbol{k})$ and $b_{\lambda}^{\dagger}(\boldsymbol{k})$ are annihilation and creation operators with standard commutation relations:
\begin{eqnarray} \label{comm} 
\left[b_{\lambda}(\boldsymbol{k}), b_{\lambda '}^{\dagger}({\boldsymbol{k}}^{\prime})
\right] = {\delta}^3
(\boldsymbol{k}-{\boldsymbol{k}}^{\prime})\delta_{\lambda \lambda '}\, , \quad \left[ 
b_{\lambda }(\boldsymbol{k}), b_{\lambda '}({\boldsymbol{k}}^{\prime})\right]
= \left[b_{\lambda }^{\dagger}(\boldsymbol{k}),
b_{\lambda '}^{\dagger}({\boldsymbol{k}}^{\prime})\right] = 0 \ .
\end{eqnarray}
As only the background dynamics are relevant here, we have $\phi=\phi(\eta)$, so we can directly write $f(\eta)$. Writing the Fourier amplitude in (\ref{FourierA}) as ${\cal A}(\eta ,k)\equiv f(\eta) a(\eta) A(\eta,k)$ we get the most convenient form of the equation of motion
\begin{align} \label{eomfourier}
{\cal A}''(\eta ,k)+\left(k^2-\frac{f''}{f}\right){\cal A}(\eta ,k)=0 \ .
\end{align}
The commutation relation between $A_i$ and the canonical momentum fixes the normalisation as
\begin{align} \label{wronskian}
  \mA(\eta,k){ \mA'}^{*}(\eta,k) - { \mA'}(\eta,k)\mA^{*}(\eta,k)=i \ .
\end{align}

\para{The energy-momentum tensor.}

The electric and magnetic fields are
\begin{equation}
E_{\mu}=F_{\mu \nu} u^{\nu} = ( 0 , - a^{-1} {A}'_i ) \, ,\quad B_{\mu }=\frac12 \epsilon _{\mu \nu \a\b} u^\b F^{\nu \a} \, ,
\end{equation}
where the four-velocity is $u^{\mu}=a^{-1}\left(1, \boldsymbol{0}\right)$. The electromagnetic energy-momentum tensor is
\begin{equation} \label{stresstensor}
T_{\mu \nu} = f^2(\phi)\bigg(F_{\mu \a }{F_{\nu}}^\a - \frac{1}{4}g_{\mu \nu}F_{\a \beta}F^{\a \beta}\bigg)\, ,
\end{equation}
so the energy density is
\begin{align} \label{rho}
  u^\a u^\b T_{\a\b} &= - \f{f^2(\phi)}{2a^2}A'_i{A'}^i-\f{f^2(\phi)}{4}F_{ij}F^{ij} =-\f{f^2(\phi)}{2}\bigg(E_iE^i+B_iB^i\bigg)\nonumber \\&\equiv u^\a u^\b T_{E,\a\b} + u^\a u^\b T_{B,\a\b} \ .
\end{align}
Using \Eqs{FourierA}, \re{complet} and \re{comm} we get
\bea
\label{rhoB} 
\rho_B(\eta) &\equiv& - \av{u^\a u^\b T_{B,\a\b}} 
= \int_{0}^{\infty}\f{{\rm d}k}{2\pi ^2}
\left[\frac{k}{a(\eta)}\right]^4\left \vert {\cal A}(\eta,k)\right\vert^2 \equiv \int_{0}^{\infty}\f{\rmd k}{k} \d_B^2(\eta,k)\\
  \label{rhoE} \rho_E(\eta) &\equiv& - \av{u^\a u^\b T_{E,\a\b}} 
  = f^2(\phi)\displaystyle\int_{0}^{\infty}\f{{\rm d}k}{2\pi ^2}
\left[\frac{k}{a(\eta)}\right]^4k^{-2} \left\vert \left[\f{ {\cal A}(\eta,k)}{f(\phi)}\right]' \right\vert^2 
\equiv \int_{0}^{\infty}\f{\rmd k}{k} \d_E^2(\eta,k) ,
\nonumber \\&
\eea
where $\av{}$ denotes vacuum expectation value and we have defined the magnetic and electric power spectra
\bea
  \label{deltaB} \delta^2_B(\eta,k) &\equiv& \f{k^5}{2\pi^2} \f{1}{a(\eta)^4} \left\vert {\cal A}(\eta,k)\right\vert^2 \\
  \label{deltaE} \delta^2_E(\eta,k) &\equiv& \f{k^3}{2\pi^2} \f{f(\phi)^2}{a(\eta)^4} \left\vert \left[\f{ {\cal A}(\eta,k)}{f(\phi)}\right]'\right\vert^2 \ .
\eea

\subsection{IR-UV matching} \label{sec:match}

The equation of motion \re{eomfourier} does not have a general analytical solution. Solutions have been considered for specific forms of $f(\eta)$ (or $f(\phi)$) \cite{Ratra:1991bn, Demozzi:2009fu, Ferreira:2013sqa, Bonvin:2013tba, Ferreira:2014hma}. In this work we want to make as few assumptions about $f(\eta)$ as possible. We therefore solve \Eq{eomfourier} separately in the infrared (IR) and the ultraviolet (UV) regimes, \ie in the sub- and super-Hubble limits respectively, and patch the solutions together, as done in \Ref{Demozzi:2009fu}.

In the large-momentum (UV) limit we get an oscillating solution. Taking the positive frequency mode (corresponding to the Bunch--Davies vacuum) with the normalisation (\ref{wronskian}) gives
\begin{equation} \label{solUV}
  \mA_{\rm UV}(\eta,k) = \f{1}{\sqrt{2k}} e^{-ik\eta} \ .
\end{equation}
In the small momentum (IR) limit, \Eq{eomfourier} can be written as
\begin{equation}
\left\lbrace f^2(\eta)\left[\f{\mA_{\rm IR}(\eta,k)}{f(\eta)}\right]'\right\rbrace'=0 \, ,
\end{equation}
with the general solution \cite{Demozzi:2009fu}
\bea \label{solIR}
  \mA_{\rm IR}(\eta, k) &=& C_1(k)f(\eta) + C_2(k) f(\eta) \int^\eta_{\eta_i} \f{\rmd\tau} {f(\tau)^2} 
  \el &\equiv& 
  C_1(k) f(\eta) + C_2(k) F(\eta) \, ,
\eea
where $\eta_i$ is some initial time and the second line defines the quantity $F$ [changing $\eta_i$ just corresponds to a redefinition of $C_1(k)$].

We consider slow-roll inflation and work to zeroth order in the slow-roll parameters, which corresponds to exponential expansion $a\propto e^{Ht}$, where $t$ is cosmic time and $H$ is constant; in terms of conformal time, we have $a=-1/(\eta H)$. We normalise the scale factor so that $a_0=1$, where the subscript 0 refers to today. We now impose the condition that the UV and IR solutions \re{solUV} and \re{solIR} and their first derivatives match at $aH=-\eta^{-1}=\se k$, where $\se$ is a constant that is not much different from unity. As the IR and UV solutions are valid in the well-separated regions, $k\ll aH$ and $k\gg aH$, respectively, there is a range of possible choices of where to match them, and $\se$ parametrises the related uncertainty. The matching assumes that the form of $f$ is such that modes that have crossed into the IR do not cross back into the UV. As we will consider upper bounds on the magnetic field amplitude, this is a conservative assumption. If a mode were to cross back from the IR into the UV, its amplitude would oscillate and not grow, so the magnetic field amplitude would decrease. The matching conditions are
\bea \label{match}
  \mA_{\rm UV}[-(\se k)^{-1} , k] = \mA_{\rm IR}[-(\se k)^{-1} , k] \, , \quad \mA_{\rm UV}'[-(\se k)^{-1} , k] = \mA_{\rm IR}'[-(\se k)^{-1} , k] . 
  \quad\quad
\eea
Inserting \Eqs{solUV} and \re{solIR} into \Eq{match} and taking into account the normalisation condition \re{wronskian}, we get the unique (up to a phase) solution
\bea \label{C}
  \sqrt{2k} C_1(k) &=& F'[-(\se k)^{-1}] + i k F[-(\se k)^{-1}] \el
  \sqrt{2k} C_2(k) &=& - f'[-(\se k)^{-1}] - i k f[-(\se k)^{-1}] \ .
\eea
We expect this approximation to capture the leading super-Hubble term.

\subsection{Approximate versus exact solution} \label{sec:exact}

Let us check the accuracy of the matching solution in a case where the exact solution is known.
We consider the coupling function $f(\eta)=(\eta/\eta_\rme)^{-\a}\propto a^\a$, where $\a$ is a constant and the subscript ``$\rme$'' refers to the end of inflation. 
\para{Exact solution in the power-law case.}

In this case \Eq{eomfourier} reduces to a Bessel equation, with the solutions \cite{Ratra:1991bn, Ferreira:2013sqa}
\begin{equation} \label{solex}
  {\cal A}(\eta ,k) = \sqrt{-k\eta}\left[N_1(k)H^{(1)}_{\a+\ha}(-k\eta)+N_2(k)H^{(2)}_{\a+\ha}(-k\eta)\right] \, ,
\end{equation}
where $H^{(i)}_{\a+\ha}$ are Hankel functions and $N_i(k)$ are integration constants. In the UV limit $k\rightarrow\infty$ we have
\begin{equation}
  H^{(1)}_{\a+\ha}(-k\eta) \simeq \sqrt{\f{2}{\pi (-k\eta)}} e^{-i \f{\pi}{2} \left(\a+\ha\right)} e^{-ik\eta} \, ,
\end{equation}
so the normalisation (\ref{wronskian}) and the property $H^{(2)}_{\a+\ha}(-k\eta)=H^{(1)}_{\a+\ha}(-k\eta)^*$ fix $N_1(k)=\sqrt{\pi/(4k)}$ and $N_2(k)=0$. The IR limit of the properly normalised mode is then (up to a constant phase)
\bea \label{Aex}
  {\cal A}(\eta ,k) &=& \sqrt{\f{-\eta}{4\pi}} \left[ e^{-i \f{\pi}{2} \left(\a+\ha\right)} 2^{\a+\ha} \Gamma\left(\a+\ha\right) (-k\eta)^{-\a-\ha} \right. \\
  && \left. + e^{i\f{\pi}{2}(\a+\ha)} 2^{-\a-\ha}  \Gamma\left(-\a-\ha\right) (-k\eta)^{\a+\ha} \right] \ .
\eea
For $\a+\ha>0$ the first term of \Eq{Aex} dominates, and the magnetic amplitude is, from \Eq{deltaB},
\bea \label{magex1}
  \d_B(\eta,k) &=& \frac{2^{\a-1}}{\pi^{\frac{3}{2}}} \Gamma\left(\a+\ha\right) \frac{(-\eta)^{-\a}}{a^2(\eta)} k^{-\a+2} \ .
\eea
For $\a+\ha<0$ the second term is dominant, and we get
\bea \label{magex2}
  \d_B(\eta,k) &=& \frac{2^{-\a-2}}{\pi^{\frac{3}{2}}} \Gamma\left(-\a-\ha\right) \frac{(-\eta)^{\a+1}}{a^2(\eta)} k^{\a+3} \ .
\eea

\para{Matched solution in the power-law case.}

Let us compare these exact results to the approximate solution obtained by the matching procedure. With a suitable choice of $\eta_i$ we have $F=\eta^{\a+1}\eta_\rme^{-\a}/(2\a+1)$, so the general IR solution \re{solIR} is 
\begin{equation}
  \mA_{\rm IR}(\eta,k) = C_1(k) \left( \frac{\eta}{\eta_\rme} \right)^{-\a} + \f{C_2(k)}{2\a+1} \eta_\rme \left( \frac{\eta}{\eta_\rme} \right)^{\a+1} \, ,
\end{equation}
and the matching conditions \re{C} give
\begin{align} \label{Cpower}
\begin{dcases}
  \sqrt{2k} C_1(k) = \f{\a + 1 - i \se^{-1}}{2\a+1} (-\eta_\rme)^{-\a} (\se k)^{-\a} \\ 
  \sqrt{2k} C_2(k) = - ( \a + i \se^{-1} ) (-\eta_\rme)^{\a} (\se k)^{\a+1} 
\end{dcases}
 \, ,
\end{align}
so the solution is
\bea \label{Apower}
  {\cal A}_{\rm IR}(\eta,k) &=& \frac{1}{2\a+1} \sqrt{\f{-\eta}{2}} \left[ ( \a + 1 - i \se^{-1} ) \se^{-\a} (-k\eta)^{-\a-\ha} \right. \el
  && \left. + ( \a + i \se^{-1} ) \se^{\a+1} (-k\eta)^{\a+\ha} \right] \ .
\eea
The magnetic amplitude is, for $\a+\ha>0$,
\bea \label{mag1}
  \d_B(\eta,k) &=& \frac{\sqrt{(\a+1)^2 + \se^{-2}}}{2\pi (2\a+1)} \se^{-\a} \frac{(-\eta)^{-\a}}{a(\eta)^2} k^{-\a+2} \, ,
\eea
and for $\a+\ha<0$ we get
\bea \label{mag2}
  \d_B(\eta,k) &=& \frac{\sqrt{\a^2 + \se^{-2}}}{2 \pi |2\a+1|} \se^{\a+1} \frac{(-\eta)^{\a+1}}{a(\eta)^2} k^{\a+3} \ .
\eea

The approximate solution \re{Apower} has the same dependence on $k$ and $\eta$ as the exact mode \re{Aex} in the super-Hubble limit, so the magnetic field spectrum is qualitatively correct. The sub-leading terms of \Eq{solex} are not captured by the matching approximation, so in the case $\a+\ha>0$ we do not get a right estimate of the electric power spectrum (as the leading term of \Eq{deltaE} then vanishes). However, we will find that in this case the amplitude of the magnetic power spectrum is anyway too small to match observations, so this does not limit our results.
The numerical prefactor of the amplitude depends on $\se$, but for reasonable choices it is close to the exact result. For the often studied case of a scale-invariant spectrum, $\a=2$ or $\a=-3$, the ratio of the approximate and exact result is $\frac{\sqrt{9+\se^{-2}}}{15}\se^{-2}$; for matching at Hubble crossing, $\se=1$, this factor is $\approx0.2$, and
the correct amplitude is obtained for $\se\approx0.5$.

\subsection{Requirements for successful magnetogenesis}

\para{Strong coupling, backreaction and perturbations.}

In order to obtain successful magnetogenesis by breaking conformal invariance with the function $f$, some well-known conditions have to be satisfied. The first one is that for the model to stay perturbative, the electromagnetic coupling constant has to remain small \cite{Ratra:1991bn, Demozzi:2009fu}. When we rescale the vector potential as $A^\a\rightarrow f^{-1}A^\a$ to obtain a canonically normalised kinetic term, the fine structure constant $\a_{\mathrm{EM}}$ scales as $\a_{\mathrm{EM}}\rightarrow f^{-2}\a_{\mathrm{EM}}$. At late times we have to recover standard electromagnetism, so then $f=1$ and $\a_{\mathrm{EM}}^{-1}\approx137$ (neglecting the running). A common assumption in the literature is that $f\geq1$, but since radiative corrections are proportional to $f^{-2}\a_{\mathrm{EM}}$, it could be allowed to be slightly smaller than unity while still marginally maintaining perturbativity. In order to be conservative, we impose the limit $f\geq\fmin$, where $\fmin\geq0.1$ quantifies the dependence on the assumed lower limit. We assume that $f=1$ at the end of inflation and after.

The second condition is that in order for the calculation in a fixed inflationary background to be consistent, the energy density in the electric and magnetic fields has to be negligible during inflation \cite{Demozzi:2009fu, Fujita:2012rb, Ferreira:2013sqa}. Significant electromagnetic contribution does not necessarily spoil inflationary magnetogenesis, but its influence on inflationary dynamics has to be taken into account \cite{Kanno:2009ei, Kanno:2008gn,Dimopoulos:2010xq,Dimopoulos:2011ym, Karciauskas:2016pxn}.
The third condition is that the electromagnetic perturbations must not be so large as to spoil the success of the inflationary mechanism of generating the observed scalar perturbation \cite{Giovannini:2007aq, Suyama:2012wh, Giovannini:2013rme, Ringeval:2013hfa, Bonvin:2013tba, Fujita:2014sna, Ferreira:2014hma, Ade:2015cva}. We discuss the backreaction and perturbation constraints in \sec{sec:BR}.

It is well known that the coupling function $f\propto\eta^{-\a}$ leads to either the strong coupling problem, the backreaction problem, or to a too small amplitude \cite{Demozzi:2009fu, Fujita:2012rb}; in \Ref{Ferreira:2013sqa} a more complicated form of $f$ was proposed to get around these problems. We now show that, for slow-roll inflation, producing a power-law spectrum with an amplitude consistent with observations, while avoiding both the strong coupling and backreaction problems, would be very contrived regardless of the form of $f$.

\section{Constraints on magnetogenesis} \label{sec:con}

\subsection{Theoretical and observed power spectra} \label{sec:theobs}

\para{Theoretical power spectrum.} 

We assume that the magnetic power spectrum in the range where magnetic fields have been observed is a power law, $\delta_B^2(\eta_0,k)\propto k^{n_B+3}$. We denote the time that corresponds to the beginning of inflation by $\eta_\rmb=-\se^{-1} k_\rmb^{-1}$, and the times that correspond to matching the longest and shortest wavelength modes in the observable region by $\eta_1=-\se^{-1} k_1^{-1}$ and $\eta_2=-\se^{-1} k_2^{-1}$ respectively. It is convenient to choose (with no loss of generality) the integration constant $\eta_i=\eta_2$ in \Eq{solIR}. The magnetic field amplitude is then, from \Eqs{deltaB} and \re{solIR},
\bea \label{theo}
  \delta_B(\eta, k) &=& \f{k^2}{2\pi a(\eta)^2} \left| \sqrt{2 k} C_1(k) f(\eta) + \sqrt{2 k} C_2(k) f(\eta) \int^\eta_{\eta_2} \f{\rmd\tau} {f(\tau)^2} \right| \ .
\eea
The fact that $\delta_B$ is a power law does not imply that $f$ is a power law. However, in appendix \ref{sec:app} we show that $f$ can deviate significantly from a power law only for a small range of e-folds, which would not affect our conclusions. Therefore we take $f$ to be a power law in the region where the observed modes are generated. For $\eta_1<\eta<\eta_2$, we then have $f=D(\eta/\eta_\rme)^{-\a}$, where $D$ is a constant. We make no assumptions about the form of $f$ outside the observable range, except that $f\geq\fmin$ throughout and $f(\eta_\rme)=1$. We will not be able to write the solution down fully even for the modes in the observable range, because the second term in \Eq{theo} depends on the history of the function $f(\eta)$ everywhere between the beginning of the observable region and the end of inflation. At the end of inflation the $C_1(k)$ term in \Eq{theo} is given by
\bea
  \sqrt{2 k} C_1(k) f(\eta_\rme) &=& F'[-(\se k)^{-1}] + i k F[-(\se k)^{-1}] \el
  &=& \frac{D^{-1}}{2\a+1} \left[ \left(\a+1-i\se^{-1}\right) \left( \frac{k}{k_\rme} \right)^{-\a} 
 \right. \el && \left.
  + \left(\a+i\se^{-1}\right) \left( \frac{k_2}{k_\rme} \right)^{-2\a-1} \left( \frac{k}{k_\rme} \right)^{\a+1} \right] \, ,
\eea
where $k_\rme=-(\se\eta_\rme)^{-1}$. This term involves the values of $F$ only during the time the modes are generated. In contrast, the second term depends also on the value of $F$ after the observable modes have left the Hubble radius, and thus on the entire history of $f$ after $\eta_1$. At the end of inflation the $C_2(k)$ term is given by
\bea
  \sqrt{2 k} C_2(k) F(\eta_\rme) &=& - \left\lbrace f'[-(\se k)^{-1}] + i k f[-(\se k)^{-1}] \right\rbrace F(\eta_\rme) \el
  &=& D \left(\a+i\se^{-1}\right) \eta_\rme^{-1} \left( - \eta_\rme \se k \right)^{\a+1} \int_{\eta_2}^{\eta_\rme} d\tau f(\tau)^{-2} \el
  &\equiv& D \left(\a+i\se^{-1}\right) \eta_\rme^{-1} \left( - \eta_\rme \se k \right)^{\a+1} \kappa (\eta_\rme-\eta_2) \el
  &\approx& - \kappa D \left(\a+i\se^{-1}\right) \left( \frac{k_2}{k_\rme} \right)^{-1} \left( \frac{k}{k_\rme} \right)^{\a+1} \, ,
\eea
where on the last line we have taken into account that the universe expands by a large factor between the end of the observational region and the end of inflation, so $\eta_\rme/\eta_2\ll1$. We have introduced the constant $\kappa\equiv (\eta_\rme-\eta_2)^{-1}\int_{\eta_2}^{\eta_\rme} d\tau f(\tau)^{-2}$, which parametrises the lack of knowledge about the form of $f$. We have $f\geq\fmin$, so $0<\kappa<\fmin^{-2}\leq100$.

Adding the two contributions, we get
\begin{align}
\label{AIR}
  \sqrt{2 k} \mA(\eta_\rme,k) &= \frac{D^{-1}}{2\a+1} \left\lbrace \left(\a+1-i\se^{-1}\right) \left( \frac{k}{k_\rme} \right)^{-\a} 
 \right.  \el & \left.
   + (\a+i\se^{-1}) \left[ 1 - (2\a+1) \kappa f(\eta_2)^2 \right] \left( \frac{k_2}{k_\rme} \right)^{-2\a-1} \left( \frac{k}{k_\rme} \right)^{\a+1} \right\rbrace \, .
\end{align}
As the matching approximation is expected to capture only the leading super-Hubble modes, the first term should dominate for $\a+\ha>0$, so
\bea \label{dpos}
  \d_B(\eta_\rme,k) = \f{\sqrt{(\a+1)^2+\se^{-2}} D^{-1}}{2\pi (2\a+1) a_\rme^2} \left( \frac{k}{k_\rme} \right)^{2-\a} k_\rme^2 \, ,
\eea
where $a_\rme\equiv a(\eta_\rme)$. For $\a+\ha<0$, the second term should dominate, so
\begin{align}\label{dneg}
  \d_B(\eta_\rme,k) = \f{\sqrt{\a^2+\se^{-2}} D^{-1}}{2\pi |2\a+1| a_\rme^2} \left[ 1+|2\a+1|\kappa f(\eta_2)^2 \right] \left( \frac{k_2}{k_\rme} \right)^{-2\a-1} \left( \frac{k}{k_\rme} \right)^{\a+3} k_\rme^2 \ .
\end{align}
There is a potential problem in that the coefficient of \Eq{dneg} can be large. The ratio of \Eqs{dneg} and \re{dpos} is $\sim |1 - (2\a+1)\kappa f(\eta_2)^2| (k/k_2)^{2\a+1}$. At $k_2$, the ratio is independent of the spectral index. In the case $\a+\ha>0$, the consistency of our treatment therefore requires that the terms 1 and $-(2\a+1)\kappa f(\eta_2)^2$ accurately cancel, whereas in the case $\a+\ha<0$ we must have $|2\a+1|\kappa f(\eta_2)^2\gg1$ (this is what happens in the power-law case discussed in \sec{sec:exact}). However, when we compare to observations, we will see that the term \re{dpos} is anyway negligibly small compared to the observed magnetic field amplitude, and unless the condition $|2\a+1|\kappa f(\eta_2)^2\gg1$ is satisfied, the second term is too small as well.

\para{The observed power spectrum.}

\begin{figure*}
\begin{center}
\includegraphics[]{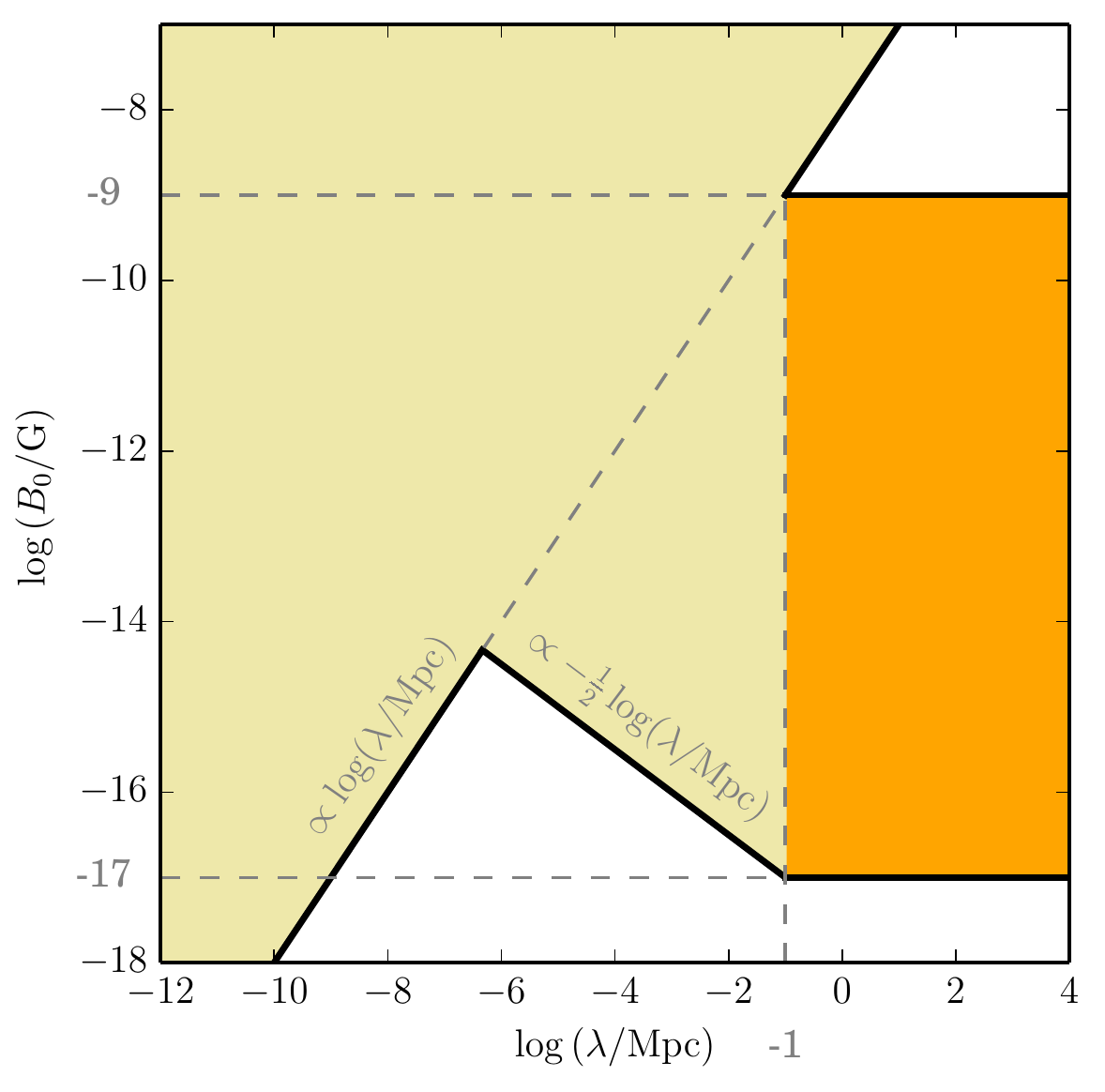}
\caption{Constraints on $B_0$, the magnetic field strength today, as a function of the comoving scale $\lambda$. White regions are observationally excluded. Values above the diagonal line $\propto \log(\lambda/\mathrm{Mpc})$ have been processed by MHD and cannot be simply related to the primordial values. Orange marks the region that we use for constraints.}
\label{fig:obs}
\end{center}
\end{figure*}

Observational constraints on the amplitude of magnetic fields on large scales today are summarised in \Ref{Durrer:2013pga}. Combining theoretical bounds from magnetohydrodynamical turbulence decay, observational limits from Faraday rotation measurement, cosmic microwave background (CMB) anisotropies and spectral distortions, gamma ray observations, ultra-high-energy cosmic ray observations and constraints from initial seed fields for galactic dynamo, they get the following constraints for the current magnetic field strength $B_0$ in units of gauss (G) as a function of scale $\l$:
\begin{align}
\log\bigg(\frac{B_0}{\mathrm{G}}\bigg) \in
 \begin{cases}
   \left[-17,-9\right] & \text{if } \log\left(\frac{\lambda}{\mathrm{Mpc}}\right)\in\left[-1,4\right]  \vspace{0.1cm}\\
   \left[ -17.5-\frac{1}{2} \log\left(\frac{\lambda}{\mathrm{Mpc}}\right),-8+\log\left(\frac{\lambda}{\mathrm{Mpc}}\right) \right]     & \text{if } \log\left(\frac{\lambda}{\mathrm{Mpc}}\right)\in\left[-6.3,-1\right] \vspace{0.1cm} \\ 
      \left[-8+\log\left(\frac{\lambda}{\mathrm{Mpc}}\right), \text{no upper limit} \right] & \text{if } \log\left(\frac{\lambda}{\mathrm{Mpc}}\right)\in [-9,-6.3]
  \end{cases}
  \ .
\end{align}
These constraints on the observed power spectrum are shown in \fig{fig:obs}. The observational upper limit $B_0\lesssim 10^{-9}$ G comes mainly from the CMB and the Faraday rotation of radio emission spectra of distant quasars.
The most important constraint is that on comoving wavelengths $\l>0.1$ Mpc, the amplitude has to be between $10^{-17}$ G and $10^{-9}$ G today. The lower limit comes from the non-observation of inverse Compton scattering from very high energy gamma rays and has the caveat that this could possibly be explained by plasma instabilities instead of large-scale coherent magnetic fields \cite{Durrer:2013pga}. We follow the common interpretation of the observations in terms of a magnetic field.
The Planck collaboration has reported the constraint $B_0<4.4$ nG at $\l=$ 1 Mpc \cite{Ade:2015cva}\footnote{Note that the conventions of the Planck team differ from ours. We take $\d_B(\eta_0,k)$ as the estimate of the magnetic field amplitude on length scale $2\pi/k$. The value $B_\l$ quoted by the Planck team is, in our notation, $\ha (2\pi)^{-(n_B+3)/2} \Gamma(\frac{n_B+4}{2})^{1/2} \d_B(\eta_0,2\pi/\l)$.}. They also get the marginalised constraint $n_B<-0.008$, but this has to be interpreted with care, as the constraint for a fixed amplitude can be quite different. The smaller the amplitude, the more freedom there is for the spectral index. Because the Planck data only gives an upper bound on $B_0$, it is consistent with any value of the spectral index for a small enough amplitude.
The tilt $n_B$ of a power-law solution corresponds to the slope in \fig{fig:obs} and the allowed values between 0.1 and $10^4$ Mpc translate to the upper and lower limits $|n_B+3|<3.2$, \ie $-6.2<n_B<0.2$.

On small scales, the evolution is affected by the non-linear coupling between magnetic fields and plasma in the early universe. The system enters a turbulent regime, where energy is transferred from large to small scales and is eventually dissipated into heating up the plasma \cite{Durrer:2013pga}. 
The line ${\rm log} (B_0/\mathrm{G}) \propto {\rm log} (\lambda/$Mpc) in \fig{fig:obs} corresponds to the largest possible regions that can have been processed by causal magnetohydrodynamics. Magnetic fields initially above the line (when scaled as $B_0 = B_{\rm in} a_{\rm in}^2, \lambda = \lambda_{\rm in}/a_{\rm in}$, where ``$\rm{in}$'' refers to the initial value; recall that $a_0=1$) decay through turbulence until they hit the line. Therefore there are no reliable constraints on the amplitude of primordial magnetic fields above the line.
Although there is a lower limit on the observed magnetic fields between 
and $10^{-9}$ and $0.1$ Mpc that is below the magnetohydrodynamical line, it is not clear whether the origin of these fields is primordial, and, if so, whether their amplitude is simply scaled by $1/a^2$. We therefore exclude them from the analysis, and only consider the wavelengths between $\l_2=0.1$ Mpc and $\l_1=10^4$ Mpc. Including the smaller-scale modes would significantly tighten our constraints.
On scales between $0.1$ Mpc and $10^4$ Mpc, the observational constraints on the amplitudes are below the magnetohydrodynamic line, so we assume that the magnetic fields retain their primordial spectrum and are only diluted by the expansion of the universe as $B\propto a^{-2}$. The magnetic power spectrum today is then simply related to the value at the end of inflation as $\d_{B}(\eta_0,k)=a_\rme^2 \d_{B}(\eta_\rme,k)$.

\paragraph{Observational limits and the theoretical power spectrum.}

Let us first consider the case \mbox{$\a+\ha>0$}.  From \Eq{dpos} we have
\bea \label{poslimit}
  \d_B(\eta_0,k) &=& a_\rme^2 \d_{B}(\eta_\rme,k) \el
  &=& \f{\sqrt{(\a+1)^2+\se^{-2}}}{2\pi (2\a+1)} D^{-1} \left( \frac{k}{k_\rme} \right)^{-\a+2} k_\rme^2 
  =\f{\sqrt{(\a+1)^2+\se^{-2}}}{2\pi (2\a+1)} f(-k^{-1})^{-1} k^2 \el
  &<& 7.0 \times 10^{-57} \sqrt{(\a+1)^2+\se^{-2}} \fmin^{-1} (\l/\mathrm{Mpc})^{-2} \ \mathrm{G} \, ,
\eea
where we have put in $k=2\pi/\l$ and written $\l^{-2}=2.0\times10^{-57} (\l/$Mpc$)^{-2}$ G. We have also taken into account that for $\d_B\propto k^{-\a+2}$ the constraint $|n_B+3|<3.2$ implies $0.4<\a<3.6$. For the maximum comoving wavelength on which we have observational constraints, $\l_1=10^4$ Mpc, and the extreme value $\fmin^{-1}=10$, we get $\d_B(\eta_0,k_1)<7\times10^{-64} \sqrt{(\a+1)^2+\se^{-2}}$ G, many orders of magnitude below $10^{-17}$ G for any reasonable value of $\se$. The case \mbox{$\a+\ha>0$} is therefore excluded.

Let us now look at the case $\a+\ha<0$. As $\d_B\propto k^{\a+3}$, the constraint $|n_B+3|<3.2$ gives $-4.6<\a<-1.4$. We get the maximum amplitude when the second term in the square brackets in \Eq{dneg} is much larger than 1, in which case
\bea \label{neglimit}
  \d_B(\eta_0,k) &=& a_\rme^2 \d_B(\eta_\rme,k) 
  =\f{\sqrt{\a^2+\se^{-2}}}{2\pi} \kappa f(\eta_2) \left( \frac{k}{k_2} \right)^{\a+3} k_2^{2} \ .
\eea
The amplitude is enhanced by the factor $f(\eta_2)$. At $k_2=2\pi/(0.1$ Mpc) we have  $\d_B(\eta_0,k_2)<10^{-54}\sqrt{\a^2+\se^{-2}}\kappa f(\eta_2)$ G, so we need $\sqrt{\a^2+\se^{-2}}\kappa f(\eta_2)\geq10^{37}$ to reach $10^{-17}$ G (as $\a<0$, $f(\eta_2)>1$ guarantees that there is no strong coupling anywhere in the region of interest). However, $f(\eta_2)$ cannot be increased without limit, as the electromagnetic energy density must not become so large as to spoil the success of inflation.

\subsection{Backreaction and perturbations} \label{sec:BR}

\paragraph{Background constraint from backreaction.}

We have to check that the contribution of the electromagnetic field to the energy density, \Eqs{rhoB} and \re{rhoE}, disturbs neither the background evolution during inflation nor the generation of the curvature perturbation. Let us first consider the background contribution. A conservative limit is that the electromagnetic contribution to the energy density is less than the total energy density during inflation, $\rho_{BE}\equiv\rho_B+\rho_E<3 \Mpl^2 H^2$. Note that if the averaged equation of state during reheating is larger than $1/3$, the electromagnetic energy density could overtake the inflaton decay products and dominate the radiation density, in contradiction with observations. We do not consider the constraints arising from avoiding that.

The energy density of the electromagnetic field is given by an integral over the modes generated from the beginning of inflation. From \Eqs{rhoE}, \re{deltaE}, \re{solIR} and \re{C} we get for the electric contribution
\bea
  \rho_E(\eta) &>& \f{1}{a(\eta)^4 f^2(\eta)} \f{1}{2\pi^2} \int_{k_
\rmb}^{-\se^{-1}\eta^{-1}} \f{\rmd k}{k} k^3 |C_2(k)|^2 \el
  &>& \f{1}{a(\eta)^4 f^2(\eta)} \f{1}{4\pi^2} \int_{k_1}^{-\se^{-1}\eta^{-1}} \rmd k k\left[ f'(-\se^{-1} k^{-1})^2 + k^2 f(-\se^{-1} k^{-1})^2 \right] \, ,
\eea
where we have dropped the contribution from the era before the observable modes are generated, since we know nothing about it. We know the form of $f$ between $k_1$ and $k_2$, but between $k_2$ and $k_\rme$ we know only the initial and final values $f(\eta_2)$ and $f(\eta_\rme)=1$, and that $f\geq\fmin$. We therefore get constraints during the time that the observable modes are generated and at the end of inflation, but not between.

During the time that the observational modes are generated, $\eta_1\leq\eta\leq\eta_2$, we obtain the following lower bound for the electric energy density
\bea \label{rhoElim}
  \rho_E(\eta) &>& \frac{\a^2+\se^{-2}}{8\pi^2 (\a+2)} \se^{-2} \left[ 1 - \left( -\eta \se k_1 \right)^{2\a+4} \right] H^4 \ .
\eea
This gives a non-trivial constraint when $\a+2<0$. For the magnetic energy density, we get similarly from \Eqs{rhoB}, \re{deltaB}, \re{solIR} and \re{C}
\bea \label{rhoBlim}
  \rho_B(\eta) &>& \frac{\a^2+\se^{-2}}{8\pi^2 (2\a+1)^2 (\a+3)} \se^{-4} \left[ 1 - \left( - \eta \se k_1 \right)^{2\a+6} \right] H^4 \, ,
\eea
where we have dropped terms that are too small to be relevant. The magnetic field contribution is subdominant to the electric field contribution, except possibly for $\a\gtrsim-2$, in which case the constraint $\rho_{BE}<3 \Mpl^2 H^2$ is anyway trivially satisfied. The background constraints are strongest at $\eta_2$.
Requiring $\rho_E<3\Mpl^2 H^2$ with \Eq{rhoElim} gives the constraint $\a>-3.0$, using\footnote{\label{footnote:Hbound}The limit on $H/\Mpl$ comes from the limit on the tensor-to-scalar ratio $r$. The inflationary tensor power spectrum is $\mathcal{P}_t=\frac{2}{\pi^2}(H/\Mpl)^2=r\mathcal{P}_\zeta$, which yields $H/\Mpl=3\times10^{-5}(\frac{r}{0.07})^{1/2}$, where the latest constraint from combined BICEP2/Keck and Planck data is $r<0.07$ \cite{Array:2015xqh}.} $H/\Mpl<3\times10^{-5}$. This value is for $\se=1$, but the dependence on $\se$ is only logarithmic.

At the end of inflation we have for the electric field
\bea
  \rho_E(\eta_\rme) &>& \f{1}{a_\rme^4} \f{1}{4\pi^2} \int_{k_1}^{k_\rme} \rmd k k\left[ f'(-\se^{-1} k^{-1})^2 + k^2 f(-\se^{-1} k^{-1})^2 \right] \el
  &=& \f{1}{a_\rme^4} \f{1}{4\pi^2} \left\lbrace \int_{k_1}^{k_2} \rmd k k\left[ f'(-\se^{-1} k^{-1})^2 + k^2 f(-\se^{-1} k^{-1})^2 \right] \right. \el
  && \left. + \int_{k_2}^{k_\rme} \rmd k k\left[ f'(-\se^{-1} k^{-1})^2 + k^2 f(-\se^{-1} k^{-1})^2 \right] \right\rbrace \el
  &>& \f{1}{a_\rme^4} \f{1}{4\pi^2} \left\lbrace \int_{k_1}^{k_2} \rmd k k\left[ f'(-\se^{-1} k^{-1})^2 + k^2 f(-\se^{-1} k^{-1})^2 \right] \right. \el
  && \left. + \int_{k_2}^{k_\rme} \rmd k k f'(-\se^{-1} k^{-1})^2 \right\rbrace \ .
\eea
In the integral from $k_1$ to $k_2$, we can insert the known form of $f$. We can write the integral from $k_2$ to $k_\rme$ as $\int_{k_2}^{k_\rme} \rmd k k f'(-\se^{-1} k^{-1})^2=4\se^{-2}\int_{x_\rme}^{x_2} \rmd x (\rmd f/\rmd x)^2\geq 4\se^{-2} [f(\eta_2)-1]^2/(x_2-x_\rme)$, where $x=(\se k)^{-4}$. The inequality follows from writing the integral as an average over the range from $x_\rme$ to $x_2$ and using the fact that variance is non-negative:
$\int_{x_\rme}^{x_2} \rmd x (\rmd f/\rmd x)^2\equiv(x_2-x_\rme)\av{(\rmd f/\rmd x)^2}\geq(x_2-x_\rme)\av{\rmd f/\rmd x}^2$.
The result is, dropping terms that are too small to be relevant,
\bea \label{rhoElime}
  \rho_E(\eta_\rme) &>& \frac{\se^{-2}}{\pi^2} f^2 (\eta_2) \left(\frac{k_2}{k_\rme} \right)^4 \left\lbrace   \frac{\a^2+\se^{-2}}{8 (\a+2)} \left[ 1 - \left( \frac{k_1}{k_2} \right)^{2\a+4} \right] + 1 \right\rbrace H^4 \ .
\eea
For the magnetic field energy density we get
\bea \label{rhoBlime}
  \rho_B(\eta_\rme) &>& \frac{\a^2+\se^{-2}}{8\pi^2 (\a+3)} \se^{-4} \kappa^2 f^2(\eta_2) \left( \frac{k_2}{k_\rme} \right)^4 \left[ 1 - \left( \frac{k_1}{k_2} \right)^{2\a+6} \right] H^4 \  .
\eea
For $\a\gtrsim-2$, the constraint from the magnetic field can be more stringent than the one from the electric field.

Making use of \Eq{neglimit} to replace $f(\eta_2)$, the limit $\rho_E(\eta_\rme)+\rho_B(\eta_\rme)<3 \Mpl^2 H^2$ can be evaluated with \Eqs{rhoElime} and \re{rhoBlime} and we obtain
\bea \label{Blim1}
  \d_B(\eta_0,k) &<& 4 \sqrt{3} \pi^2 \frac{\Mpl}{H} \kappa \left\lbrace \frac{\se^{2}}{2\a+4} \left[ 1 - \left( \frac{k_1}{k_2} \right)^{2\a+4} \right] + \frac{\kappa^{2}}{2\a+6} \left[ 1 - \left( \frac{k_1}{k_2} \right)^{2\a+6} \right] \right. \el
  && \left. + \frac{4 \se^2}{\a^2+\se^{-2}} \right\rbrace^{-\ha} \left( \frac{k}{k_2} \right)^{\a+3} ( \se^{-1} \l_\rme )^{-2} \el
  &\approx& 4.6 \times 10^{-51} \kappa \left( \frac{r}{0.07} \right)^{-\frac{1}{2}} \left\lbrace \frac{\se^{2}}{2\a+4} \left[ 1 - \left( \frac{k_1}{k_2} \right)^{2\a+4} \right] + \frac{\kappa^{2}}{2\a+6} \left[ 1 - \left( \frac{k_1}{k_2} \right)^{2\a+6} \right] \right. \el
  && \left. + \frac{4 \se^2}{\a^2+\se^{-2}} \right\rbrace^{-\ha} \left( \frac{k}{k_2} \right)^{\a+3} \left( \frac{\se^{-1} \l_\rme}{\mathrm{Mpc}} \right)^{-2} G \ .
\eea
The quantity $\se^{-1} \l_\rme$ is the smallest wavelength that exits the Hubble radius during inflation. It depends on the inflationary scale and the dynamics of preheating,
\bea \label{le}
  \se^{-1} \l_\rme &=& \frac{2\pi}{a_\rme H} 
  =\frac{2\pi \sqrt{3} \Mpl}{a_\rme \sqrt{\rho_\rme}} \el
  &=& 2\pi \sqrt{3} \Mpl \left[ \frac{g_*(T_\reh)}{g_*(T_0)} \right]^{\frac{1}{12}} \left( \rho_\reh \rho_{\c0} \right)^{-\frac{1}{4}} \left( \frac{\rho_\reh}{\rho_\rme} \right)^{\frac{1+3\bar w}{6(1+\bar w)}} \el
  &\approx& 140 \times10^{-6} \left[ \frac{g_*(T_\reh)}{10.75} \right]^{\frac{1}{12}} \left( \frac{\rho_\reh}{\rho_{\mathrm{BBN}}} \right)^{-\frac{1}{4}} \left( \frac{\rho_\reh}{\rho_\rme} \right)^{\frac{1+3\bar w}{6(1+\bar w)}} \mathrm{Mpc} \el
  &>&  140 \times10^{-6}  \left( \frac{\rho_\rme}{\rho_{\mathrm{BBN}}} \right)^{-\frac{1}{4}} \mathrm{Mpc} 
  \approx 5.2 \times 10^{-23} \left( \frac{r}{0.07} \right)^{-\frac{1}{4}} \mathrm{Mpc} \, ,
\eea
where we have used the adiabatic relation $g_{*S}(T) T^3 a^3=$ constant, ``$\reh$'' refers to reheating, $\rho_{\c0}=\frac{\pi^2}{15} T_0^4$ with $T_0=2.725$ K is the radiation energy density today, $-1/3<\bar w<1$ is the average (over the number of e-folds) equation of state between the end of inflation and the onset of the radiation era, and $g_{*}$ is the effective number of energy degrees of freedom, which we have assumed to be the same as the effective number of entropy degrees of freedom $g_{*S}$.\footnote{Conversely, taking into account that the dependence on $g_*(T_\reh)$ is weak, assuming that $g_*(T_\reh)$ is not much larger than the Standard Model maximum value 106.75, we get the upper limit $\se^{-1}\l_\rme<180$ pc. In any case, we know from observations of large-scale structure that $\se^{-1}\l_\rme\lesssim0.4$ Mpc \cite{Viel:2005qj, Viel:2013apy}, independently of the details of inflation.} On the next-to-last line we have inserted the lower limit on the reheating energy density from big bang nucleosynthesis, which is $\rho_{\mathrm{BBN}}=\frac{\pi^2}{30} g_*(T_{\mathrm{BBN}}) T_{\mathrm{BBN}}^4$ where $T_{\mathrm{BBN}}=4.7$ MeV is the lowest possible temperature for thermalisation of Standard Model particles after inflation \cite{deSalas:2015glj}; as $g_*(4.7 \ \mathrm{MeV})=10.75$, this gives $\rho_\reh^{1/4}>\rho_{\mathrm{BBN}}^{1/4}=6.4$ MeV. On the last line we have written $H/\Mpl=3\times10^{-5} (r/0.07)^{\frac{1}{2}}$, see footnote~\ref{footnote:Hbound}, corresponding to $\rho_\rme^{1/4}=3\times10^{16} (r/0.07)^{\frac{1}{4}}$ GeV.
Note that we can get the lower limit in two different ways. If $\bar w\leq1/3$, we maximise $\rho_\reh$ by putting it equal to $\rho_\rme$, in which case $\bar w$ is irrelevant. If $\bar w\geq1/3$, we minimise $\rho_\reh$ by putting it equal to $\rho_{\mathrm{BBN}}$, in which case $\bar w=1/3$ gives the minimal amplitude. In both cases the density factors reduce to $(\rho_\rme/\rho_{\mathrm{BBN}})^{-1/4}$.

Inserting \Eq{le} into \Eq{Blim1}, we have
\bea \label{eq:deltaBend:cond1}
  \d_B(\eta_0,k) &<& 1.7 \times 10^{-6} \kappa \left\lbrace \frac{\se^{2}}{2\a+4} \left[ 1 - \left( \frac{k_1}{k_2} \right)^{2\a+4} \right] + \frac{\kappa^{2}}{2\a+6} \left[ 1 - \left( \frac{k_1}{k_2} \right)^{2\a+6} \right] \right. \el
  && \left. + \frac{4 \se^2}{\a^2+\se^{-2}} \right\rbrace^{-\ha} \left( \frac{k}{k_2} \right)^{\a+3} \mathrm{G} \ .
\eea
Note that the dependence on $r$ (\ie on the inflationary energy scale) drops out. In the case where the second term (\ie the magnetic contribution) or the last term dominates, the constraint is anyway too weak to be relevant, so only the  first term is important. Taking into account $\a\geq-3.0$, the most difficult amplitude to reproduce is the one on the largest wavelengths. We get the following $\a$-dependent constraint on the amplitude:
\bea \label{eq:deltaBend:cond2}
  \d_B(\eta_0,k_1) &<& 1.7 \times 10^{-6} \se^{-1} \kappa \left\lbrace \frac{1}{2\a+4} \left[ 1 - \left( \frac{k_1}{k_2} \right)^{2\a+4} \right] \right\rbrace^{-\ha} \left( \frac{k_1}{k_2} \right)^{\a+3} \mathrm{G} \ .
\eea
The maximum value is reached for $\a=-3$,
\bea
  \d_B(\eta_0,k_1) &<& 2 \times 10^{-11} \se^{-1} \kappa \mathrm{G} \, ,
\eea
and we get the minimal value for $\a=-1.4$,
\bea
  \d_B(\eta_0,k_1) &<& 2 \times 10^{-14} \se^{-1} \kappa \mathrm{G} \ .
\eea

While in principle we could get a magnetic field amplitude of $10^{-9}$ G by taking $\kappa=100$, this would correspond to an extremely fine-tuned situation.
The function $f$ (and thus the QED coupling) would have to jump immediately at the end of the observational window to the value $\fmin$ at the limit of perturbativity, and then to the standard value $f=1$ immediately before the end of inflation. Also, in this case the contribution from $f'^2$ between $k_2$ and $k_\rme$ (which we have neglected) would be large, and we would have to jointly minimise the contribution from $\kappa$ (which calls for a rapid shift in $f$) and the constraint from $f'^2$ (which calls for $f$ not to change rapidly). We do not consider such optimisation, but now consider the limit from perturbations, which turns out to be stronger than the background limit.

\para{Constraint from perturbations.}

We consider the gauge invariant curvature perturbation $\zeta = -\psi -H\delta\rho/\dot{\rho}$, which receives contributions both from the inflaton and the electromagnetic field. Assuming single-field slow-roll inflation, the spectrum of the inflaton contribution $\zeta_{I}$ is
\bea \label{PR}
  \mathcal P_{\zeta_I} &=& \frac{1}{8 \pi^2 \epsilon} \frac{H^2}{\Mpl^2} \, ,
\eea
where $\epsilon$ is the first slow-roll parameter. The contribution of the electromagnetic energy density to the curvature perturbation is
\bea
\label{PBE}
  \zeta_{BE} &=& -H \frac{\d \rho_{BE}}{\dot\rho}
   =  \frac{\delta\rho_{BE}}{6 \epsilon \Mpl^2 H^2} 
   =   \frac{4 \pi^2}{3} \mathcal{P}_{\zeta_I} \frac{\delta\rho_{BE}}{H^4} \, ,
\eea
where we have inserted $\dot\rho=-6\epsilon\Mpl^2 H^3$ and used \Eq{PR}.

The component $\zeta_{BE}$ is non-Gaussian since  $\rho_{BE}$ given in \Eq{rho} is quadratic in the vector potential ${\cal A}$. The electromagnetic field is therefore constrained both by the observed amplitude of the total curvature perturbation ${\cal P}_{\zeta} = 2.2\times 10^{-9}$ \cite{Ade:2015xua} and by observational bounds on primordial non-Gaussianity. Detailed investigation of the non-Gaussian signatures~\cite{Caprini:2009vk, Caldwell:2011ra, Motta:2012rn, Jain:2012ga, Barnaby:2012tk, Lyth:2013kah, Fujita:2013pgp, Nurmi:2013gpa, Ferreira:2014hma} is beyond the scope of our current work. To get a rough constraint, we approximate the magnitude of the bispectrum as $\langle\zeta^3\rangle \sim {\cal P}_{\zeta_{BE}}^{\frac{3}{2}}$ and use the Planck limit on local non-Gaussianity $f_{\rm NL} = 2.5\pm 5.7$ \cite{Ade:2015ava}. This constrains the contribution of ${\cal P}_{\zeta_{BE}}$ to the total spectrum to be at most at the percent level. Moreover, if the magnetic contribution ${\cal P}_{\zeta_{BE}}$ is strongly scale dependent, the measured scale invariance of primordial perturbations yields a quantitatively similar constraint. To account for both constraints, we parameterise the electromagnetic contribution to the power spectrum as $\mathcal{P}_{\zeta_{BE}}\leq10^{-2}\mathcal{P}_{\zeta}\sigma_\zeta$, with the default value $\sigma_\zeta=1$. If $\mathcal{P}_{\zeta_{BE}}$ is close to scale invariant, its contribution could possibly be larger, given that the electromagnetic and matter perturbations could equilibrate in reheating, so that there are no isocurvature perturbations observable in the CMB. The parameter $\sigma_\zeta$ can be adjusted to suit, and our limits anyway turn out not to depend strongly on $\sigma_\zeta$.

As the electromagnetic field is assumed to be energetically subdominant during inflation, its fluctuations generate isocurvature perturbations. Therefore the total curvature perturbation $\zeta = \zeta_I +\zeta_{BE}$ is not conserved but evolves during and after inflation. The evolution of $\zeta$ depends on the details of reheating, see e.g. \Ref{Bonvin:2011dt}. After reheating the universe contains a large number of charged particles that rapidly dissipate the electric field. Because the magnetic fields scale as radiation $B^2\propto a^{-4}$ on superhorizon scales, the magnetic contribution to the curvature perturbation (\ref{PBE}) remains constant during radiation domination. A detailed analysis of the evolution is beyond the scope of our current work, and we look only at the electromagnetic contribution to the curvature perturbation at the end of inflation, $\zeta_{BE}(\eta_{\rm end})$. If its contribution to the total power spectrum is small compared to the observed amplitude, the electromagnetic contribution to the final curvature perturbation is expected to be small.  

Using the expression (\ref{rho}) of the electromagnetic energy density, we can  express the spectrum of $\zeta_{BE}$ at the end of inflation as 
\baq
\label{PBE_step1}
\nonumber
  {\cal P}_{\zeta_{BE}}^{\rm end} &=&  \left(\mathcal{P}^{\rm end}_{\zeta_I}\right)^2\; \frac{(2\pi)^5 k^3}{9 H^8}\int\rmd^3 q
\left\lbrace 
\left[P_{E}(q)P_{E}(|{\bf k-q}|)+P_{B}(q)P_{B}(|{\bf k-q}|)\right] 
 \right. \\&& \times \left.
  \left(1+{\rm cos}^2\theta \right) - 4{\rm cos}\theta\; {\rm Re}\left[ P_{BE}(q)P_{BE}(|{\bf k-q}|)\right]
\right\rbrace
 \el  &\gtrsim& \left(\mathcal{P}^{\rm end}_{\zeta_I}\right)^2\; \frac{(2\pi)^5 k^3}{9 H^8}\int_{k_1< q,|{\bf k}-{\bf q}| < k_2} \rmd^3 q
\left\lbrace
\left[P_{E}(q)P_{E}(|{\bf k-q}|) + P_{B}(q)P_{B}(|{\bf k-q}|)\right] 
 \right. \el && \times \left.
  \left(1+{\rm cos}^2\theta \right)- 4{\rm cos}\theta\; {\rm Re}\left[ P_{BE}(q)P_{BE}(|{\bf k-q}|)\right] \right\rbrace \, ,
\eaq 
where $\theta$ is the angle between ${\bf q}$ and ${\bf k-q}$. The magnetic and electric power spectra used here are related to \Eqs{deltaB} and (\ref{deltaE}) as $P_{B}\equiv\delta_B^2/(4\pi k^3 f^2)$ and $P_{E}\equiv\delta_E^2/(4\pi k^3 f^2)$. The cross spectrum is defined as $\langle B_i{}^*({\bf k})E_j{}({\bf k'})\rangle = - (2\pi)^3 \delta ({\bf k}-{\bf k'}) i \epsilon_{ijk} \hat{k}^k P_{BE}(k)$, with $P_{BE} \equiv k ({\cal A}/f) ({\cal A}/f)'{}^*/[a^4 (2\pi)^3]$. On the second line we have taken into account that while we cannot evaluate the convolution integrals in \Eq{PBE_step1} without knowing the full form of the coupling $f(\eta)$, we can get a lower limit for the full result by calculating the result in the range $k_1<k<k_2$. As the third term in the integrand is not positive definite, this limit is not watertight. However, as the electric contribution usually dominates over the magnetic contribution and the electric-magnetic cross term is oscillatory, we expect that our lower limit is valid, barring fine-tuned cases.
We use the power-law form $f = D(k_{\rm end}/k)^{-\alpha}$ in the window $k_1<k<k_2$ and substitute the asymptotic solution (\ref{AIR}), which leaves us with convolutions of the form $\int\rmd^3 q q^n |{\bf k}-{\bf q}|^n$. We use methods similar to the ones presented in \Refs{Mack:2001gc, Caprini:2003vc, Kahniashvili:2005xe} and estimate the integrals by including contributions from the three regimes, $q\ll k$, $q\sim k$ and $q\gg k$. This gives a reasonable order of magnitude estimate, which suffices for our purposes.
\bea
\label{PBE_lowerbound}
  \frac{\mathcal{P}_{\zeta_{BE}}^{\rm end}}{ (\mathcal{P}^{\rm end}_{\zeta_I})^2} &\gtrsim& \frac{f(\eta_2)^4(\alpha^2+\se^{-2})^2}{18} \se^{-4} \left( \frac{k_2}{k_{\rm end}}\right)^{8} \left(\frac{k}{k_2}\right)^{3} \Bigg\lbrace \kappa^4 \se^{-4} I_{2\alpha+3}(k) + I_{2\alpha+1}(k) 
 \nonumber \\ & & \left.
  + \frac{4 \kappa^2 \se^{-2}}{4\alpha+7} \left[ 1 - \left( \frac{k}{k_2}\right)^{4\a+7} \right] \right\rbrace 
  \nonumber \\ &=&
   \frac{\delta^4_B(\eta_0,k) (\se^{-1}\l_\rme)^8}{18 (2\pi)^4} \left( \frac{k}{k_2} \right)^{-4\a-9} \Bigg\lbrace I_{2\alpha+3}(k) + \kappa^{-4} \se^{4} I_{2\alpha+1}(k)
  \nonumber \\ & & \left.
  + \frac{4 \kappa^{-2} \se^{2}}{4\alpha+7} \left[ 1 - \left( \frac{k}{k_2}\right)^{4\a+7} \right] \right\rbrace \, ,
\eea
where  
\beq
I_n(k) \equiv \frac{1}{2n+3} \left[ 1 - \left(\frac{k}{k_2}\right)^{2n+3} \right] + \frac{2}{n+3} \left( \frac{k}{k_2}\right)^{n} \left[ \left(\frac{k}{k_2}\right)^{n+3} - \left( \frac{k_1}{k_2} \right)^{n+3} \right] \, ,
\eeq
and on the second line of \Eq{PBE_lowerbound} we have used \Eq{neglimit}.

The condition $\mathcal{P}_{\zeta_{BE}}^{\rm end}\leq10^{-2} \sigma_\zeta\mathcal{P}_{\zeta}^{\rm end}$ ensuring that the spectrum of $\zeta_{BE}$ at the end of inflation is smaller than the observed spectrum of curvature perturbations then translates \Eq{PBE_lowerbound} into an upper bound on $\delta_B(\eta_0,k)$. Violation of this condition does not strictly rule out the setup, because the curvature perturbation evolves over reheating and this might decrease $\zeta$ to the observed level. However, we may at least argue that CMB observations imply quite non-trivial constraints for such scenarios. 

Given $-3.0<\a<-1.4$, the middle term in \Eq{PBE_lowerbound} (which arises from the purely electric contribution) gives the strongest bound, which comes from the longest wavelength. Using $\mathcal{P}_{\zeta_{BE}}^{\rm end}\leq10^{-2} \sigma_\zeta\mathcal{P}_{\zeta}^{\rm end}$, assuming that the electromagnetic contribution is clearly subdominant, $\mathcal{P}_{\zeta_{I}}\approx\mathcal{P}_{\zeta}=2.2\times 10^{-9}$, and using the lower limit \re{le} on $\se^{-1}\l_\rme$, the constraint from \Eq{PBE_lowerbound} is 
\bea \label{pertbound}
  \delta_B(\eta_0,k_1) < 5 \times10^{-15} \abs{4\a+5}^{1/4} \left( \frac{r}{0.07} \right)^{\frac{1}{2}} \sigma_k^{-1} \sigma_\zeta^{1/4} \kappa \mathrm{G} \ .
\eea
The dependence on $\a$ is weak, since $\abs{4\a+5}^{1/4}\simeq 0.9$ for $\a=-1.4$ and $\simeq 1.6$ for $\a=-3$.

Note that the perturbation constraint (\ref{pertbound}) depends crucially on the assumption that the observed curvature perturbation is dominantly generated by the inflaton. It does not apply if the inflaton contribution to the total curvature perturbation is small, as in the curvaton scenario \cite{Enqvist:2001zp, Lyth:2001nq, Moroi:2001ct}. The electromagnetic energy density goes down as $a^{-4}$ after reheating, the same way as the inflaton decay products, so the curvaton will dampen its contribution to the total perturbation in the same way as it dampens the inflaton contribution. Hence the perturbative constraint disappears in this case.

\section{Conclusions} \label{sec:conc}

\para{Comparison to previous work.}

A number of earlier papers have put constraints of varying degrees of generality on magnetogenesis models where the conformal symmetry is broken by $f(\phi)$, starting from \Ref{Demozzi:2009fu}, which noted that a power-law form leads to either the backreaction or strong coupling problem, or to a too small amplitude. In \Ref{Ferreira:2013sqa} the authors consider a form of $f(\phi)$ tuned to avoid these problems, as well as the effect of optimising the reheating history (and assuming the curvaton scenario to decouple the amplitude of inflationary perturbations and the Hubble scale), with a maximum amplitude of $10^{-17}$ G around $10^4$ Mpc today and $10^{-13}$ G on Mpc scales. They consider the strong coupling problem and backreaction problem, but not the effect of the electromagnetic perturbations, which gives our strongest limit. These values are consistent with our background-only limit, although this need not have been the case, as the spectrum of \Ref{Ferreira:2013sqa} is not a power law. In \Ref{Ferreira:2014hma} the authors consider a power-law form for $f$ and take into account the bispectrum (which we did not consider). They find a maximum value of $10^{-15}$ G on Mpc scales. Contrary to our case where $f$ can have arbitrary form between the end of the observational window and the end of inflation, both papers find the maximum amplitude for the lowest inflationary scale.

Some studies have derived upper limits by adopting a power-law description of quantities related directly to the magnetic field rather than to the coupling function and other more model-independent studies \cite{Suyama:2012wh, Ringeval:2013hfa, Fujita:2014sna}.\footnote{The paper \cite{Green:2015fss} also aims at model-independent bounds. However, when going from their Eq.~(3.10) to their Eq.~(3.11), they assume that the limit $\int_{k_1}^{k_2} \frac{\rmd k}{k} \delta^2_E(\eta,k)\ll1$ implies $\delta^2_E(\eta,k)\ll1$ for $k_1<k<k_2$. 
This is a strong restriction on $\delta^2_E(\eta,k)$ and consequently on the coupling function $f(\phi)$.} However, in this case it is not possible to consider the important strong coupling constraint. The study \cite{Fujita:2012rb} is rather model-independent and takes the strong coupling issue into account, but it only gives an upper limit on the minimum value of the magnetic field, not on the maximum possible value.

It is easy to understand why our background constraint is independent of the inflationary scale. The maximum magnetic energy density is proportional to the total energy density during inflation. If the inflationary scale is higher, then on the one hand the magnetic field amplitude at the end of inflation is larger. On the other hand, the universe has to expand by a larger factor to end up at the radiation energy density today, fixed by the measured CMB temperature. For instantaneous reheating, the energy density in the inflaton decay products goes like $\propto a^{-4}$, like the magnetic energy density, so the two effects exactly compensate each other. Taking into account different reheating histories and the change in the number of relativistic degrees of freedom does not change the conclusion for the maximum magnetic field amplitude. For the perturbative constraints, the amplitude of the induced curvature perturbations is proportional to the magnetic energy density, but the total curvature perturbation, to which we compare, is fixed by observation, so the first above mentioned effect is absent, and we are left with the dependence $\d_B(\eta_0,k)\propto r^{1/2}$. If we used the theoretical inflaton spectrum \re{PR} instead and assumed $\epsilon<1$ (as done in \Ref{Fujita:2014sna}) or fixed $\epsilon=0.01$ (as done in \Ref{Ferreira:2014hma}), our perturbative limit would have a less stringent dependence on $r$.

\para{Summary.}

We find that the requirement that the power spectrum today on the observed scales is a power law determines the coupling $f$ to essentially be a power law, $f\propto\eta^{-\a}$. Crucially, we make no assumptions about the form of $f$ outside the window where the observable modes are generated, apart from avoiding strong coupling. We take this range of scales to be between 0.1 Mpc and $10^4$ Mpc. From the constraint on the backreaction on the background evolution in the observable window and at the end of inflation, optimising over all possible reheating histories described by an average equation of state $-1/3  <\bar{w} < 1$, we find the limit
\bea
\label{eq:summary:background}
  \d_B(\eta_0, k_1) &<& 2 \times 10^{-14} \se^{-1} \kappa\, \mathrm{G} \, ,
\eea
where $\se$ parametrises the uncertainty due to our approximation of matching the sub- and super-Hubble modes when solving for the magnetic field, and is not expected to be much different from unity. The parameter $\kappa$ parametrises the unknown form of $f$ outside the observational window, and is $<100$. This is a conservative bound, and it is not guaranteed that there exists a function $f$ that can saturate it. If such a function exists, it will be extremely fine-tuned, and we would naturally expect $\kappa\ll1$.
We also find that the spectral index of the magnetic field has to lie between $-3<n_B<0.2$, \ie $-1.4<\a<-3$.

Taking into account that the electromagnetic perturbations cannot disturb the power spectrum of curvature perturbations too much, we get the limit 
\bea
\label{eq:summary:pert}
  \delta_B(\eta_0, k_1) < 5 \times10^{-15} \abs{4\a+5}^{1/4} \left( \frac{r}{0.07} \right)^{1/2} \sigma_k^{-1} \sigma_\zeta^{1/4} \kappa\, \mathrm{G} \, ,
\eea
where $\sigma_\zeta$ is the maximum fractional contribution of the electromagnetic power spectrum contribution to the total curvature power spectrum, with $\sigma_\zeta=1$ corresponding to a maximum of $10^{-2}$. This limit is more stringent, but it can be avoided if the curvature perturbation is not generated by the inflaton but for example via the curvaton mechanism. We only consider the perturbative limit at the end of inflation, but evolution during and after inflation could modify this bound.

The assumption of a power-law form for the observed magnetic field and the lever arm of five orders of magnitude in wavelength are important for our limits. If we extend the constraints down to lengths of $\lambda_2=10^{-6}$ Mpc, the requirement on the amplitude tightens linearly with $\lambda_2$, which makes the electromagnetic contribution to the curvature perturbations too large if the latter are generated through the standard inflationary mechanism. If we go all the way to $10^{-9}$ Mpc, the backreaction becomes too large at the level of the background already and the mechanism we discussed is ruled out altogether. 
Considering a more general spectrum could loosen the constraints, and more detailed investigation of the bispectrum and other higher-order statistics could lead to more stringent constraints.

\acknowledgments

We thank Ruth Durrer, Francesc Ferrer, Till Sawala and Martin Sloth for correspondence. The research leading to these results has received funding from the European Research Council under the European Union's Horizon 2020 program (ERC Grant Agreement No. 648680), and from the Science and Technology Facilities Council (STFC Grants No. ST/K00090X/1 and No. ST/L005573/1).

\appendix

\section{Form of the coupling function} \label{sec:app}

\para{From the magnetic power spectrum to the coupling function.}

Here we show that if the magnetic spectrum is a power law, $\d_B(\eta_0,k)^2\propto k^{n_B+3}$ in the observable region $k_1\leq k\leq k_2$, then $f$ can significantly deviate from a power law only for a small range of e-folds. From \Eqs{C} and \re{theo} we have
\bea
    \d_B(\eta_\rme, k)^2 &\propto& k^5 \left| C_1(k) + C_2(k) \int^{\eta_\rme}_{\eta_2} \f{\rmd\tau} {f(\tau)^2} \right|^2 
   \el&\propto& 
k^4 \left. \left[ g'(\eta)^2 + k^2 g(\eta)^2 \right] \right|_{\eta=-(\se k)^{-1}} \, , \quad
\eea
where we have denoted $g(\eta)\equiv - F(\eta) - \kappa (\eta_2-\eta_\rme) f(\eta)>0$. Writing $\dot g\equiv \rmd g/\rmd N=\eta_\rme e^N g'$ and $n \equiv \frac{1}{2} (n_B - 3)$,\footnote{According to the observational constraints discussed in \sec{sec:con}, $-6.2<n_B<0.2$, so $n<0$.} we get
\bea \label{geq}
\mathrm{e}^{- 2 n N} \propto \dot{g}^2 + \se^{-2} g^2 \ .
\eea
Writing $g(N) = h(N) \mathrm{e}^{ - n N } $, the solution of \Eq{geq} is given by
\bea \label{gsin}
  h &=& h_0 \sin[\phi(N)] \\
  \dot h - n h &=& h_0 \se^{-1} \cos[\phi(N)] \, ,
\eea
where $h_0>0$ is a constant. The condition $g>0$ implies $0<\phi<\pi$. Eliminating $h$, we find that $\phi(N)$ satisfies the equation
\bea \label{phidot}
  \dot\phi = \se^{-1} + n \tan\phi \ .
\eea

\para{Sum of two power-laws.}

If $\dot\phi=0$, the constant $\phi\equiv\phi_c$ is given by $\se^{-1} + n \tan\phi_c=0$. Then $g$ is a power-law, $F + \kappa (\eta_2-\eta_\rme) f\propto e^{ - n N }\propto\eta^{-n}$. This implies that
\bea
  f(\eta) = D \left( \frac{\eta}{\eta_2} \right)^{-n} + \frac{1}{2} \left[ \sqrt{D^2 - \frac{4}{(2 n + 1) \kappa (1-\eta_\rme/\eta_2)}} -D \right] \left( \frac{\eta}{\eta_2} \right)^{n+1} \, ,
\eea
where $D$ is a constant. Unless $D$ is zero or tuned to make the second term vanish, $f$ is not a power law. However, in order to get a large enough amplitude to match observations, $D$ must be very large (for the reasons discussed in \sec{sec:theobs}), so the second term is negligible, and $f$ is close to a power-law.

\para{Deviations from power-law behaviour.} 

If $\dot\phi\neq0$, the solution has two branches,
\bea \label{gbranches}
  \label{branch1} \se^{-1} ( \se^2 n^2 +1 ) ( N + N_c ) &=& \phi + \se n \ln( \cos\phi + \se n \sin\phi ) \\
  \label{branch2} \se^{-1} ( \se^2 n^2 +1 ) ( N + N_c ) &=& \phi + \se n \ln( - \cos\phi - \se n \sin\phi ) \, ,
\eea
where $N_c$ is an integration constant. The first branch covers $0<\phi<\phi_c<\pi/2$ and the second covers $\phi_c<\phi<\pi$.
If $\phi$ does not go near $0$ or $\pi$, the modification to the power-law amplitude due to $\sin\phi$ is less than unity, and occurs only for a limited range of e-folds, so $f$ is essentially a power law. 
If $\phi$ approaches $0$ or $\pi$, the amplitude of $f$ can be damped by an arbitrarily large factor at small or large $N$, respectively (note that, by construction, the observed power spectrum is unaffected, as the decrease in $f$ is compensated by an increase in $f'$). However, this only changes $f$ in a small region that has to be tuned to be near either the end or the beginning of the observational window. Even in this fine-tuned case, $f$ deviates from a power law only for a small range of e-folds, and this does not change our results.

\bibliographystyle{JHEP}
\bibliography{mag}

\end{document}